\documentclass[preprint,showpacs,superscriptaddress,nofootinbib]{revtex4}
  \usepackage{graphicx,subfigure}
  \begin{document}
  \title{Probing Spectator Scattering and Annihilation Corrections\\
   in $B_{s}$ $\to$ $PV$ Decays}
  \author{Qin Chang}
  \affiliation{Institute of Particle and Nuclear Physics,
               Henan Normal University, Xinxiang 453007, China}
  \affiliation{State Key Laboratory of Theoretical Physics,
               Institute of Theoretical Physics,
               Chinese Academy of Sciences, Beijing, China}
  \author{Xiaohui Hu}
  \affiliation{Institute of Particle and Nuclear Physics,
              Henan Normal University, Xinxiang 453007, China}
  \author{Junfeng Sun}
  \affiliation{Institute of Particle and Nuclear Physics,
              Henan Normal University, Xinxiang 453007, China}
  \author{Yueling Yang}
  \affiliation{Institute of Particle and Nuclear Physics,
              Henan Normal University, Xinxiang 453007, China}

  \begin{abstract}
  Motivated by the recent LHCb measurements on $\bar{B}_{s}$ $\to$ $\pi^{-}K^{*+}$
  and $\bar{B}_{s}$ $\to$ $K^{\pm}K^{*\mp}$ decay modes,
  we revisit the $B_{s}$ $\to$ $PV$ decays within QCD factorization framework.
  The effects of hard-spectator scattering and annihilation corrections are studied in detail.
  After performing a $\chi^2$-fit on the end-point parameters $X_A^{i,f}$ ($\rho_A^{i,f}$,
  $\phi_A^{i,f}$) and $X_H$ ($\rho_H$, $\phi_H$) with available data,
  it is found that although some possible mismatches exist, the universalities of
  $X_A^{i,f}$ and $X_H$ in $B_s$ and $B_{u,d}$ systems are still allowed within
  theoretical uncertainties and experimental errors.
  With the end-point parameters gotten from $B_{u,d}$ $\to$ $PV$ decays,
  the numerical results and detailed analyses for the observables of
  $\bar{B}_{s}$ ${\to}$ $\pi K^{\ast}$, $\rho K$, $\pi\rho$, $\pi\phi$ and $K\phi$
  decay modes are presented. In addition, we have identified a few useful observables, especially the ones of $\bar{B}_{s}$ $\to$
  $\pi^{0}\phi$ decay for instance,
  for probing hard-spectator scattering and annihilation contributions.
  \end{abstract}
  \maketitle

  \section{introduction}
  \label{sec01}
  In the past years, many experimental efforts have devoted to precisely measuring
  nonleptonic two-body B meson decays, which provide a fertile ground to explore
  the underling mechanism of hadron weak decays.
  For instance, many $B_{u,d}$ decay modes with branching fractions
  $\gtrsim{\cal O}(10^{-6})$ are well measured by the B factories BABAR and Belle,
  and branching ratios of some $B_{s}$ decay modes are given with relatively
  high precision by CDF collaboration at Fermilab.
  Moreover, with the running of the Large Hadron Collider (LHC) and the upgrading
  Super-B factory Belle-II, not only the measurements on $B_{u,d}$ decays will be
  greatly refined, but also the $B_{s}$ decays are expected to be measured with
  extraordinary precision in the near future.
  Recently, using the data sample corresponding to $1.0 fb^{-1}$ of $pp$ collision,
  the $\bar{B}_s^0$ $\to$ $K^{\pm}K^{*\mp}$ and $\pi^-K^{*+}$ decays are
  firstly observed by LHCb collaboration, and their branching fractions
  are \cite{LHCbbskstak}
  \begin{eqnarray}
  {\cal B}(\bar{B}_s^0 \to K^{\pm} K^{*\mp})
  &=&
  (12.7\pm 1.9\pm 1.9) \times 10^{-6}
  \label{lhcbBskkstar}, \\
  {\cal B}(\bar{B}_s^0 \to \pi^{-}K^{*+})
   &=&
  (3.3\pm 1.1\pm 0.5) \times 10^{-6}
  \label{lhcbBspikstar},
  \end{eqnarray}
  with the total significances of $7.8\sigma$ and $3.4\sigma$, respectively.
  They are the first measurements of $\bar{B}_s^0$ meson decays to
  charmless $PV$ final states, where $P$ and $V$ denote  the lightest
  pseudoscalar and vector $SU(3)$ meson, respectively.

  Theoretically, based on the QCD factorization (QCDF) approach \cite{Beneke1},
  the perturbative QCD (pQCD) approach \cite{KLS} and
  the soft-collinear effective theory (SCET) \cite{scet},
  the latest predictions of such two quantities read (in the units of
  $10^{-6}$)\footnotemark[1]
  \footnotetext[1]{In Eqs. (\ref{qcdf}) and (\ref{pqcd}), the numerical
  results of ${\cal B}(\bar{B}_s^0 \to K^{\pm}K^{*\mp})$ mean
  ${\cal B}(\bar{B}_s^0 \to K^{*-}K^{+})$ $+$
  ${\cal B}(\bar{B}_s^0 \to K^{*+}K^{-})$ \cite{Cheng2,Ali1}.}
  \begin{eqnarray}
  \text{QCDF \cite{Cheng2}}
  & &
  \left\{\begin{array}{l}
  {\cal B}(\bar{B}_s^0 \to K^{\pm}K^{*\mp})
  =
  (10.3^{+3.0+4.8}_{-2.2-4.2}+11.3^{+7.0+8.1}_{-3.5-5.1}),
  \\
  {\cal B}(\bar{B}_s^0 \to \pi^{-}K^{*+})
  =
  7.8^{+0.4+0.5}_{-0.7-0.7};
  \end{array}\right.
  \label{qcdf} \\
  \text{pQCD \cite{Ali1}}
  & &
  \left\{\begin{array}{l}
  {\cal B}(\bar{B}_s^0 \to K^{\pm}K^{*\mp})
  =
  (6.0^{+1.7+1.7+0.7}_{-1.5-1.2-0.3}+4.7^{+1.1+2.5+0.0}_{-0.8-1.4-0.0}),
  \\
  {\cal B}(\bar{B}_s^0 \to \pi^{-}K^{*+})
  =
  7.6^{+2.9+0.4+0.5}_{-2.2-0.5-0.3};
  \end{array}\right.
  \label{pqcd} \\
  \text{SCET \cite{lu1}}
  & &
  \left\{\begin{array}{l}
  {\cal B}(\bar{B}_s^0 \to K^{\pm}K^{*\mp})
  =
  18.2^{+6.3+3.3}_{-5.0-2.7},\quad 19.7^{+5.0+2.6}_{-4.2-2.2},
  \\
  {\cal B}(\bar{B}_s^0 \to \pi^{-}K^{*+})
  =
  5.9^{+0.5+0.5}_{-0.5-0.5},\quad 6.6^{+0.2+0.7}_{-0.1-0.7}.
  \end{array}\right.
  \label{scet}
  \end{eqnarray}

  It is obvious that
  (1) For the branching ratio ${\cal B}(\bar{B}_s^0\to \pi^{-}K^{*+})$,
  all of the theoretical results agree with each other, but are larger than the LHCb data.
  (2) For the observable ${\cal B}(\bar{B}_s^0\to K^{\pm}K^{*\mp})$,
  the pQCD's predictions are consistent with the LHCb's measurement within
  errors, while the default results of QCDF and SCET are about $3.3\sigma$
  and $2.6\sigma$ larger than LHCb data, respectively.
  Here, it should be noted that the QCDF's results in Eq.(\ref{qcdf})
  correspond to a set of specified phenomenological annihilation parameters
  $(\rho_A, \phi_A)$ $=$ $(0.90, -65^{\circ})$ for $B_s$ $\to$ $VP$ decays
  and $(0.85, -30^{\circ})$ for $B_s$ $\to$ $PV$ decays \cite{Cheng2}.
  It is well known that for the $\bar{B}_s^0$ $\to$ $K^{\pm}K^{*\mp}$ decay,
  the tree amplitudes are suppressed by the CKM factor
  ${\vert}V_{ub}V_{us}^{\ast}{\vert}$ ${\sim}$ ${\cal O}({\lambda}^{4})$,
  while the penguin amplitudes proportional to the CKM factor
  ${\vert}V_{tb}V_{ts}^{\ast}{\vert}$ ${\sim}$ ${\cal O}({\lambda}^{2})$
  are usually sensitive to nonfactorizable corrections and effects from new
  interactions beyond the standard model.
  It will be very interesting to scrutinize the above possible mismatches.
  Before searching for possible solutions from new physics,
  it is essential to examine carefully whether the disagreement could be
  accommodated within the standard model.
  In this paper, the effects of weak annihilation (WA) and
  hard spectator scattering (HSS) corrections in
  $\bar{B}^0_s$ $\to$ $PV$ decays are studied
  in detail within the QCDF framework.

  Our paper is organized as follows. After a brief review of WA
  and HSS corrections in QCDF approach and recent researches in
  section II, we present our numerical analyses and discussions
  in section III. Our main conclusions are summarized in section IV.
  \section{HSS and WA corrections in the QCDF approach}
  \label{sec02}
  Combining the hard-scattering approach \cite{brodsky}
  with the power countering rules in the heavy quark limit
  $m_{b}$ $\gg$ ${\Lambda}_{QCD}$,
  M. Beneke {\it et al.} developed the QCDF approach to deal with
  the hadronic matrix elements for the $B$ meson decays
  based on the collinear factorization scheme and colour transparency
  hypothesis \cite{Beneke1,Beneke2}.
  Up to power corrections of order ${\Lambda}_{QCD}/m_{b}$,
  the factorization formula for $B$ decaying into two light meson
  can be written as
  \begin{eqnarray}
  {\langle}M_{1}M_{2}{\vert}O_{i}{\vert}B{\rangle}
  &=&
  \sum\limits_{j}\Big\{ F_{j}^{B{\to}M_{1}}
  {\int}dz\,T^{I}_{ij}(z){\Phi}_{M_{2}}(z)
   +(M_{1}{\leftrightarrow}M_{2})\Big\}
  \nonumber \\ & &
   +{\int}dx\,dy\,dz\,T^{II}_{i}(x,y,z)
  {\Phi}_{B}(x){\Phi}_{M_{1}}(y){\Phi}_{M_{2}}(z)
  \label{beneke},
  \end{eqnarray}
  where $F_{j}^{B{\to}M}$ denotes a $B$ ${\to}$ $M$ transition form factor,
  and ${\Phi}_{X}(z)$ is the light-cone wave functions for the two-particle
  Fock state of the participating meson $X$.
  Form factors and light-cone wave functions in Eq.(\ref{beneke}) are
  nonperturbative inputs.
  Form factors can be obtained from lattice QCD or QCD sum rules.
  Light-cone wave functions are universal and process-independent.
  Both $T^{I}(z)$ and $T^{II}(x,y,z)$ are hard scattering functions, which could
  be systematically calculable order by order with the perturbation theory in
  principle.

  When the chirally enhanced corrections to HHS amplitudes are estimated
  with the second line of QCDF formula Eq.(\ref{beneke}),
  the twist-3 contributions involve the logarithmically divergent integral
  \begin{equation}
  {\int}_{0}^{1}\frac{dt}{1-t}\,{\Phi}_{M}^{p,v}(t)=X_{H}
  \label{xhhs01},
  \end{equation}
  where the asymptotic forms of the twist-3 distribution amplitudes
  ${\Phi}_{M}^{p,v}(t)$ $=$ $1$ are employed.
  It is assumed that the endpoint singularity in Eq.(\ref{xhhs01}) contains
  many poorly known soft contributions and hence unfortunately does
  not admit a perturbative treatment within the QCDF framework.
  For an estimate, the divergent integral $X_{H}$ is phenomenologically
  parameterized as \cite{Beneke2}
  \begin{equation}
  X_{H}={\ln}\frac{m_{B}}{{\Lambda}_{h}}\,(1+{\rho}_{H}e^{i{\phi}_{H}})
  \label{xhhs02},
  \end{equation}
  with ${\Lambda}_{h}$ = 0.5 GeV.

  Moreover, it is found that although the WA
  amplitudes are power suppressed in the heavy quark limit and
  hence disappear from Eq.(\ref{beneke}),
  they are very important in practical application of the QCDF approach
  to nonleptonic $B$ decays, especially for the pure annihilation processes,
  such as $B_d$ $\to$ $K^+K^-$ and $B_s$ $\to$ $\pi^+\pi^-$ decays
  which have been observed experimentally \cite{Belleanni,LHCbanni,CDFanni}.
  The WA amplitudes are expressed in the terms of convolutions
  of ``hard scattering'' kernels with light cone wave functions to estimate
  the importance of annihilation contributions within the QCDF framework.
  A worse problem is that there is endpoint singularities even with
  twist-2 light cone distribution amplitudes.
  Similar to the case of the twist-3 hard scattering contributions
  parameterized by $X_{H}$ in Eq.(\ref{xhhs01}),
  another phenomenological quantity $X_{A}$ are introduced to
  parameterize the WA divergent endpoint integrals, i.e.
  \begin{equation}
  {\int}_{0}^{1}\frac{dt}{t}\,=X_{A}\,=
  {\ln}\frac{m_{B}}{{\Lambda}_{h}}\,(1+{\rho}_{A}e^{i{\phi}_{A}})
  \label{XA}.
  \end{equation}

  One cannot get any information on parameters ${\rho}_{H,A}$
  and ${\phi}_{H,A}$ from the QCDF approach.
  Because that the hard scattering terms arise first at order
  ${\alpha}_{s}$ and that the annihilation contributions are
  power suppressed, it is conservatively assumed that
  ${\rho}_{A,H}$ ${\le}$ 1 because too large value will give
  rise to numerically enhanced subleading contributions
  and put a question to the validity of the $1/m_{b}$
  power expansion of the QCDF approach.
  The strong phases ${\phi}_{A,H}$ can vary freely, which
  show the phenomenological importance of HSS and WA
  contributions to $CP$ asymmetric observables.
  In addition, different values of $X_{A,H}$ $({\rho}_{A,H},{\phi}_{A,H})$
  according to different types of final states $PP$, $PV$, $VP$
  and $VV$ are introduced in phenomenological investigation.
  And it is traditionally assumed that $X_{A}$ $=$ $X_{H}$ and
  they were universal for each decay types,
  which have been thoroughly discussed in
  Refs. \cite{Cheng2,Beneke2,Beneke3,Cheng1}.

  Motivated by recent measurements on pure annihilation
  $\bar{B}^0_s$ $\to$ $\pi^+\pi^-$ decay by CDF and LHCb
  collaborations \cite{LHCbanni,CDFanni},
  many more detailed studies on HSS and WA contributions
  are performed, for instance, by
  Refs. \cite{zhu1,chang1,xiao1,Gronau,Bobeth},
  and some interesting findings are presented.
  In Ref. \cite{Bobeth},
  the universality assumption of WA parameters is carefully tested,
  and some tensions in $B$ $\to$ $\pi K$, $\phi K^*$ decay modes
  with the standard model are presented.
  In Refs. \cite{zhu1,zhu2}, a ``new treatment" is proposed that
  complex parameters $X_A^i$ and $X_A^f$ coppresonding to WA topologies
  with gluon emission from initial and final states, respectively,
  should be treated independently and that the
  flavor dependence of $X_{A,H}^i$ should be carefully considered.
  Following such ``new treatment" of Refs. \cite{zhu1,zhu2},
  with available data of $B_{u,d,s}$
  $\to$ $PP$ decays, comprehensive statistical $\chi^2$ analyses on
  parameters $X_{A}^{i,f}$ and $X_H$ are performed in our previous
  works \cite{chang2,chang3}. The findings could be
  briefly summarized as that:
  (1) parameters $X^{i}_{A}$ and $X^{f}_{A}$ should be treated
  individually, as presented in Ref. \cite{zhu2},
  and the simplification $X_{H}$ $=$ $X_{A}^{i}$ is allowed by data;
  (2) The flavor dependences of $X_A^i$ are hardly to be clarified
  due to large experimental errors and theoretical uncertainties,
  which implies that the universality of $X_{A}^{i}$ in $B_{u,d}$
  and $B_s$ systems still persists.
  So, motivated by the forthcoming plenty data of $B_s$ decays
  provided by LHCb and Super-B experiments, it is worth to
  reevaluate the effects of HSS and WA corrections in $B_s$ $\to$ $PV$ decays in detail.

  \section{numerical results and discussion}
  \label{analysis}
  In the QCDF framework, the explicit expressions of the decay
  amplitudes and relevant formula of $B_s$ $\to$ $PV$ decays
  have been given in Ref. \cite{Beneke2}.
  In this paper, the $CP$-averaged branching fractions and
  $CP$ asymmetries of $\bar{B}^0_s$ $\to$ $\pi K^*$, $\rho K$,
  $\rho \pi$, $K K^*$, $\phi \pi$ and $\phi K$ decay modes
  are evaluated.
  The same definition for these observables as HFAG \cite{HFAG}
  are taken.
  The values of input parameters used in our calculations and
  analyses are summarized in Table \ref{inputs}.
  In addition, in evaluating the $CP$ asymmetry parameters
  $C_{f\bar{f}}$, $S_{f\bar{f}}$, ${\Delta}C_{f\bar{f}}$,
  ${\Delta}S_{f\bar{f}}$ and $A_{CP}^{f\bar{f}}$ of
  $\bar{B}^{0}_s$ ${\to}$ ${\pi}^{\pm}{\rho}^{\mp}$,
  $K^{\pm}K^{{\ast}\mp}$ and $K^{0}\bar{K}^{{\ast}0}(\bar{K}^{0}K^{{\ast}0})$
  decays\footnotemark[2], the conventions $\bar{f}$ $=$ ${\pi}^{+}{\rho}^{-}$,
  $K^{+}K^{{\ast}-}$ and $K^{0}\bar{K}^{{\ast}0}$ are chosen.
  \footnotetext[2]{The decay mode $\bar{B}^{0}_s$ ${\to}$
  $K^{0}\bar{K}^{{\ast}0}+\bar{K}^{0}K^{{\ast}0}$ is
  labeled as $\bar{B}^{0}_s$ ${\to}$ $K^{0}\bar{K}^{{\ast}0}(\bar{K}^{0}K^{{\ast}0})$
  for convenience.}

  \begin{table}[ht]
  \caption{The values of input parameters:
  Wolfenstein parameters, pole and running quark masses,
  decay constants, form factors and Gegenbauer moments.}
  \label{inputs}
  \begin{ruledtabular}
  \begin{tabular}{l}
  $\bar{\rho}$ $=$ $0.1453^{+0.0133}_{-0.0073}$, \quad
  $\bar{\eta}$ $=$ $0.343^{+0.011}_{-0.012}$, \quad
  $A$ $=$ $0.810^{+0.018}_{-0.024}$, \quad
  $\lambda$ $=$ $0.22548^{+0.00068}_{-0.00034}$ \cite{CKMfitter},
   \\ \hline
  $m_c$ $=$ $1.67 \pm 0.07$ GeV, \quad
  $m_b$ $=$ $4.78 \pm 0.06$ GeV, \quad
  $m_t$ $=$ $173.21 \pm 0.87$ GeV, \\
  $\frac{\bar{m}_s(\mu)}{\bar{m}_{u,d}(\mu)}$ $=$ $27.5 \pm 1.0$, \quad
  $\bar{m}_{s}(2\,{\rm GeV})$ $=$ $95 \pm 5$ MeV, \quad
  $\bar{m}_{b}(\bar{m}_{b})$ $=$ $4.18 \pm 0.03$ GeV \cite{PDG14},
  \\ \hline
  $f_{B_{s}}$ $=$ $(227.6 \pm 5.0)$ MeV, \quad
  $f_{\pi}$ $=$ $(130.41 \pm 0.20)$ MeV, \quad
  $f_{K}$ $=$ $(156.2 \pm 0.7)$  MeV \cite{PDG14,latticeaverages}, \\
  $f_{\rho}$ $=$ $(216 \pm 3)$ MeV, \quad
  $f_{K^{\ast}}$ $=$ $(220 \pm 5)$ MeV, \quad
  $f_{\phi}$ $=$ $(215 \pm 5)$ MeV, \\
  $f_{\rho}^{\bot}({\rm 1GeV})$ $=$ $(165 \pm 9)$ MeV, \quad
  $f_{K^{\ast}}^{\bot}({\rm 1GeV})$ = $(185\pm10)$ MeV, \quad
  $f_{\phi}^{\bot}({\rm1GeV})$ = $(186 \pm 9)$ MeV \cite{PBall2007},
  \\ \hline
   $A^{B_s \to \phi}_0$ = $0.26 \pm 0.06$, \quad
  $A^{B_s \to K^{\ast}}_0$ = $0.22 \pm 0.06$, \quad
  $F^{B_s \to K}_{+}$ = $0.23 \pm 0.06$ \cite{QCDSF},
  \\ \hline
  $a_{1}^{\pi}$ = 0,\quad
  $a_2^{\pi}({\rm 1\,GeV})$ = 0.25, \quad
  $a_{1}^{K}({\rm 1\,GeV})$ = 0.06, \quad
  $a_{2}^{K}({\rm 1\,GeV})$ = 0.25,\\
  $a_{1,\rho}^{||}$ = 0,\quad
  $a_{2,\rho}^{||}({\rm 1\,GeV})$  =0.15,\quad
  $a_{1,K^{\ast}}^{||}({\rm 1\,GeV})$ =0.03,\quad
  $a_{2,K^{\ast}}^{||}({\rm 1\,GeV})$ =0.11, \\
  $a_{1,\phi}^{||}$ = 0,\quad
  $a_{2,\phi}^{||}({\rm 1\,GeV})$ = 0.18 \cite{BallG}.
  \end{tabular}
  \end{ruledtabular}
  \end{table}
  \begin{figure}[ht]
  \subfigure[]{\includegraphics[width=6cm]{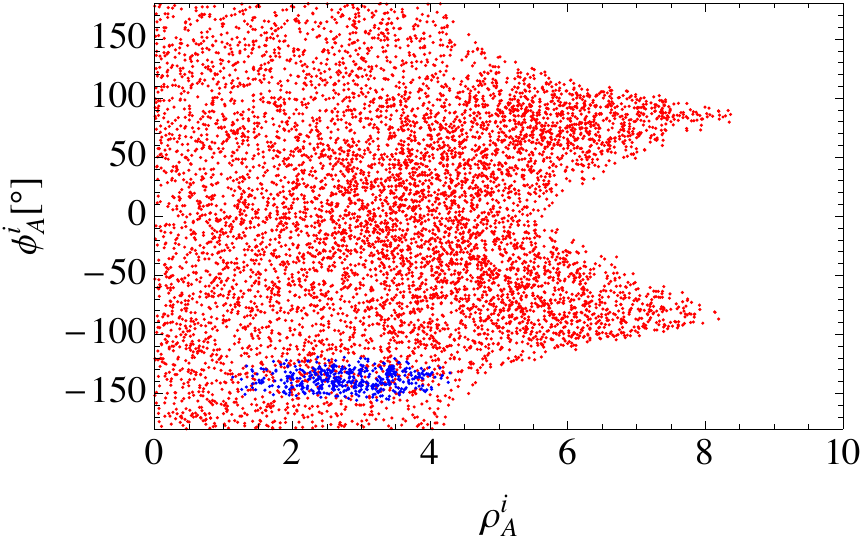}} \quad
  \subfigure[]{\includegraphics[width=6cm]{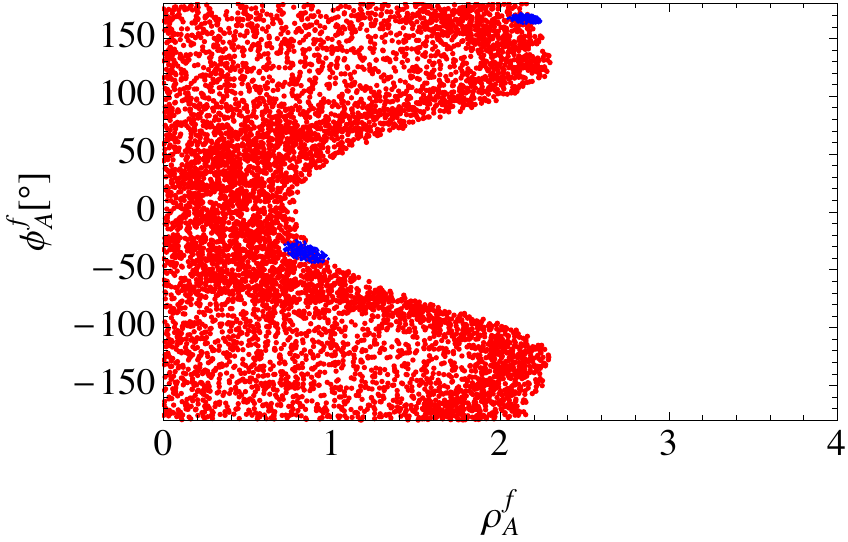}}
  \caption{The allowed spaces of parameters
  (${\rho}_{A}^{i,f}$, ${\phi}_{A}^{i,f}$) at 68\% C.L.
  with the combined constraints from
  ${\cal B}(\bar{B}_s^0 \to K^{\pm}K^{*\mp})$ and
  ${\cal B}(\bar{B}_s^0 \to \pi^{-}K^{*+})$ (red points).
  For comparison, the fitted results for $B_{u,d}$ ${\to}$ $PV$
  decays at 68\% C.L. \cite{chang4} are also shown by blue points.}
  \label{Fig1}
  \end{figure}
  \begin{figure}[ht]
  \subfigure[]{\includegraphics[width=6cm]{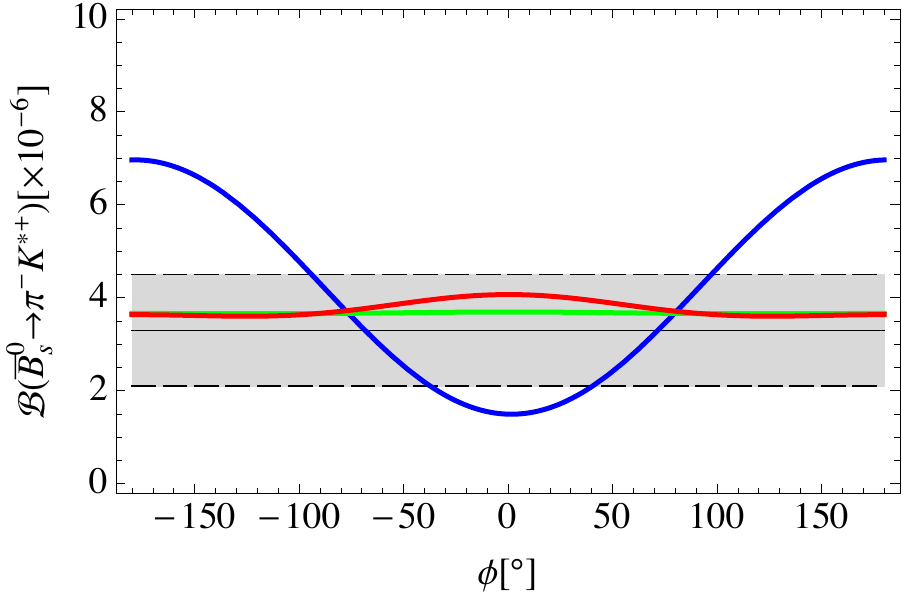}} \quad
  \subfigure[]{\includegraphics[width=6cm]{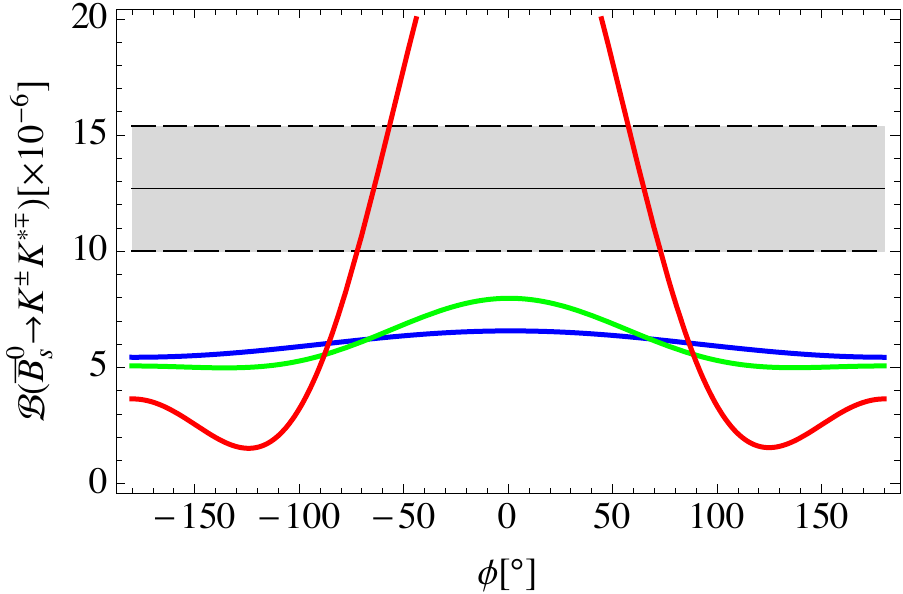}}
  \caption{The blue, green and red lines correspond to the
  dependances of ${\cal B}(\bar{B}_s^0 \to K^{\pm}K^{*\mp})$ and
  ${\cal B}(\bar{B}_s^0 \to \pi^{-}K^{*+})$ on
  $\phi_H$, ${\phi}_{A}^{i}$ and ${\phi}_{A}^{f}$ with
  $\rho_H$ = 3, ${\rho}_{A}^{i}$ = 3 and
  ${\rho}_{A}^{f}$ = 1, respectively, where
  the unconcerned parameters are set to zero.
  The shaded bands are experimental results
  with $1\sigma$ error. }
  \label{Fig2}
  \end{figure}

  Firstly, with the simplification (${\rho}_{H}$, ${\phi}_{H}$)
  = (${\rho}_{A}^i$, ${\phi}_{A}^i$) and input
  $\lambda_B$ = $0.19^{+0.09}_{-0.04}$ GeV favored by
  $B_{u,d,s}$ $\to$ $PP$ and $B_{u,d}$ $\to$ $PV$ decays
  \cite{chang3,chang4}, we perform a $\chi^2$-fit on the
  parameters (${\rho}_{A}^{i,f}$, ${\phi}_{A}^{i,f}$).
  The statistical fitting approach has been detailed in
  Appendix of Ref. \cite{chang2}.
  Using the constraints from data on
  ${\cal B}(\bar{B}_s^0 \to K^{\pm}K^{*\mp})$ and
  ${\cal B}(\bar{B}_s^0 \to \pi^{-}K^{*+})$ in
  Eqs.(\ref{lhcbBskkstar}) and (\ref{lhcbBspikstar}),
  the allowed spaces of the WA parameters at $68\%$ C.L.
  are shown in Fig.\ref{Fig1}.
  The fitted results for $B_{u,d}$ ${\to}$ $PV$ decays \cite{chang4}
  are also redisplayed for comparison.
  From  Fig.\ref{Fig1}(a), it is found that the allowed
  space of (${\rho}_{A}^{i}$, ${\phi}_{A}^{i}$) in
  $B_{u,d}$ ${\to}$ $PV$ decays entirely overlaps with
  the one in $B_{s}$ ${\to}$ $PV$ decays,
  which implies that the same parameters
  (${\rho}_{A}^{i}$, ${\phi}_{A}^{i}$)
  for $B_{u,d}$ and $B_s$ decays are still allowed by now.
  For parameters (${\rho}_{A}^{f}$, ${\phi}_{A}^{f}$) in
  $B_{s}$ ${\to}$ $PV$ decays, the red pointed region in
  Fig.\ref{Fig1}(b) shows that the allowed values of
  ${\rho}_{A}^{f}$ are strongly suppressed around
  ${\phi}_{A}^{f}$ $\sim$ $0^{\circ}$ compared with those
  with a larger ${\phi}_{A}^{f}$, which is interestingly
  similar to the situation in $B_{u,d}$ ${\to}$ $PV$
  decays shown by blue points.
  Unfortunately, due to the large theoretical uncertainties
  and only few available experimental data with large errors,
  the spaces of both (${\rho}_{A}^{i}$, ${\phi}_{A}^{i}$)
  and  (${\rho}_{A}^{f}$, ${\phi}_{A}^{f}$) in $B_s$ decays
  are hardly to be well restricted for now.

  To clarify the effects of WA and HSS corrections,
  the dependences of ${\cal B}(\bar{B}_s^0 \to K^{\pm}K^{*\mp})$
  and ${\cal B}(\bar{B}_s^0 \to \pi^{-}K^{*+})$ on parameter
  $\phi_{H,A}^{i,f}$ are presented in Fig.\ref{Fig2}.
  For the tree-dominated $\bar{B}_s^0$ $\to$ $\pi^{-}K^{*+}$ decay,
  its branching fraction is very sensitive to the HSS contributions
  related to ($\rho_H$, $\phi_H$) rather than the WA ones, as
  Fig.\ref{Fig2}(a) shows.
  So, the fitted result (red points) in Fig.\ref{Fig1}(a)
  in fact almost refers to the constraints on ($\rho_H$, $\phi_H$)
  from ${\cal B}(\bar{B}_s^0 \to \pi^{-}K^{*+})$.
  Similarly, for the penguin-dominated $\bar{B}_s^0$ $\to$
  $K^{\pm}K^{*\mp}$ decay, its branching fraction is very
  sensitive to the factorizable WA contributions related to
  (${\rho}_{A}^f$, ${\phi}_{A}^f$) rather than others as Fig.\ref{Fig2}(b) shows.
  As a result, the fitting result (red points) in Fig.\ref{Fig1}(b)
  is mainly due to the constraints from
  ${\cal B}(\bar{B}_s^0 \to K^{\pm}K^{*\mp})$.
  It is expected that the future refined measurements on
  $\bar{B}_s^0$ $\to$ $K^{\pm}K^{*\mp}$ and
  $\bar{B}_s^0$ $\to$  $\pi^{-}K^{*+}$ decays
  would perform strong constraints on factorizable
  WA and HSS corrections, respectively.

  \begin{table}[t]
  \caption{The $CP$-averaged branching fractions (in units of $10^{-6}$)
  of $\bar{B}_{s}$ ${\to}$ $\pi K^{\ast}$, $\rho K$, $\pi\rho$, $\pi\phi$,
  $K\phi$ and $KK^{\ast}$ decays.
  The first and second theoretical errors of the ``this work'' column
  are caused by uncertainties of the CKM and the other parameters in
  Table \ref{inputs}, respectively.}
   \label{br}
  \begin{ruledtabular}
  \begin{tabular}{lccccc}
  Decay Modes  & Class & This work & Cheng \cite{Cheng2} & S4 \cite{Beneke2} & FS \cite{Cheng3}
  \\ \hline
  $\bar{B}_{s}$ $\to$ $\pi^{-}K^{*+}$ & T  &$6.8^{+0.6+3.6}_{-0.6-2.8}$ & $7.8^{+0.4+0.5}_{-0.7-0.7}$      & 6.8  & $7.92 \pm 1.02$ \\
  $\bar{B}_{s}$ $\to$ $\pi^{0}K^{*0}$ & C  &$1.4^{+0.1+0.2}_{-0.1-0.2}$ & $0.89^{+0.80+0.84}_{-0.34-0.35}$ & 0.33 & $3.07 \pm 1.20$ \\
  $\bar{B}_{s}$ $\to$ $K^{+}\rho^{-}$ & T &$16.0^{+1.3+9.4}_{-1.3-7.2}$ & $14.7^{+1.4+0.9}_{-1.9-1.3}$     & 19.8 & $14.63 \pm 1.46$ \\
  $\bar{B}_{s}$ $\to$ $K^{0}\rho^{0}$ & C  &$1.4^{+0.1+0.2}_{-0.1-0.2}$ & $1.9^{+2.9+1.4}_{-0.9-0.6}$      & 0.68 & $0.56 \pm 0.24$ \\
  $\bar{B}_{s}$ $\to$ $\pi^{0}\phi$   & C, ${\rm P}_{EW}$ & $0.20^{+0.01+0.05}_{-0.02-0.04}$ & $0.12^{+0.02+0.04}_{-0.01-0.02}$ & 0.12 & $1.94 \pm 1.14$ \\
  $\bar{B}_{s}$ $\to$ $K^{0}\phi$     & P        & $0.45^{+0.02+0.07}_{-0.03-0.07}$      & $0.6^{+0.5+0.4}_{-0.2-0.3}$     &0.46&$0.41\pm0.07$\\
  $\bar{B}_{s}$ $\to$ $\pi^{0}\rho^{0}$     & WA & $0.017^{+0.001+0.001}_{-0.001-0.001}$ & $0.02^{+0.00+0.01}_{-0.00-0.01}$& 0.017 &---\\
  $\bar{B}_{s}$ $\to$ $\pi^{\pm}\rho^{\mp}$ & WA & $0.034^{+0.002+0.002}_{-0.002-0.002}$ & $0.04^{+0.00+0.02}_{-0.00-0.02}$& 0.029 &--- \\
  $\bar{B}_{s}$ $\to$ $K^{\pm}K^{*\mp}$     & P  & $17.3^{+0.7+3.1}_{-1.1-2.7}$          & $21.6^{+10.0+12.9}_{-5.7-9.3}$  & 22.7  & $16.01 \pm 1.25$ \\
  $\bar{B}_{s}$ $\to$ $K^{0}\bar{K}^{{\ast}0}(\bar{K}^{0}K^{{\ast}0})$& P  & $17.3^{+0.7+3.1}_{-1.1-2.7}$          & $20.6^{+10.9+12.8}_{-6.4-9.3}$  & 22.2  & $15.65 \pm 1.22$
  \end{tabular}
  \end{ruledtabular}
  \end{table}
  \begin{table}[ht]
  \caption{The direct and mixing-induced $CP$ asymmetries (in units of $10^{-2}$)
  of $\bar{B}_{s}$ ${\to}$ $\pi K^{\ast}$, $\rho K$, $\pi\rho$, $\pi\phi$ and
  $K\phi$ decays. The other captions are the same as Table \ref{br}.}
  \label{cp}
  \begin{ruledtabular}
  \begin{tabular}{lcccc}
  Decay Modes & This work & Cheng \cite{Cheng2} & S4 \cite{Beneke2} & FS \cite{Cheng3}
  \\ \hline
  $A_{CP}^{dir}(\bar{B}_{s}\to\pi^{-}K^{*+})$   & $-28^{+1+6}_{-1-9}$          & $-24.0^{+1.2+7.7}_{-1.5-3.9}$    & $-22.0$  & $-13.6 \pm  5.3$ \\
  $A_{CP}^{dir}(\bar{B}_{s}\to K^{+}\rho^{-})$  & $11.5^{+0.3+4.8}_{-0.5-2.8}$ & $11.7^{+3.5+10.1}_{-2.1-11.6}$   & 6.2      & $ 12.0 \pm  2.7$ \\
  $A_{CP}^{dir}(\bar{B}_{s}\to\pi^{0}K^{*0})$   & $-64^{+2+6}_{-1-6}$          & $-26.3^{+10.8+42.2}_{-10.9-36.7}$& 15.4     & $-42.3 \pm 15.8$ \\
  $A_{CP}^{dir}(\bar{B}_{s}\to\rho^{0}K^{0})$   & $51.8^{+1.4+2.7}_{-1.8-3.0}$ & $28.9^{+14.6+25.0}_{-14.5-23.7}$ & 11.6     & $-12.4 \pm 45.3$ \\
  $A_{CP}^{mix}(\bar{B}_{s}\to\rho^{0}K^{0})$   & $43^{+7+2}_{-4-2}$           & $29^{+23+16}_{-24-21}$           & ---      & $-34.8 \pm 28.5$ \\
  $A_{CP}^{dir}(\bar{B}_{s}\to\pi^{0}\phi)$     & $66^{+1+3}_{-2-5}$           & $82.2^{+10.9+9.0}_{-14.0-55.3}$  & 24.8     & $  7.3 \pm 20.1$ \\
  $A_{CP}^{mix}(\bar{B}_{s}\to\pi^{0}\phi)$     & $-73^{+1+7}_{-1-7}$          & $40^{+4+32}_{-10-53}$            & ---      & $ 43.9 \pm 17.1$ \\
  $A_{CP}^{dir}(\bar{B}_{s}\to K^{0}\phi)$      & $-2.85_{-0.12-0.73}^{+0.08+0.75}$& $-3.2^{+1.2+0.6}_{-1.4-1.3}$ & $-7.4$   & 0 \\
  $A_{CP}^{mix}(\bar{B}_{s}\to K^{0}\phi)$      & $-70^{+2+0}_{-2-0}$          & $-69^{+1+1}_{-1-1}$              & ---      & $-69.2 \pm  0.0$ \\
  $A_{CP}^{dir}(\bar{B}_{s}\to\pi^{0}\rho^{0})$ & 0 & 0 &--- &--- \\
  $A_{CP}^{mix}(\bar{B}_{s}\to\pi^{0}\rho^{0})$ & $-74^{+2+0}_{-2-0}$ & $-65^{+3+0}_{-3-0}$ &--- &---
  \end{tabular}
  \end{ruledtabular}
  \end{table}
  \begin{table}[ht]
  \caption{The $CP$ asymmetry parameters (in units of $10^{-2}$) of
  ${\bar B}^0_{s}$ $\to$ $K^{\pm}K^{*\mp}$, $\pi^{\pm}\rho^{\mp}$,
  and $K^{0}\bar{K}^{{\ast}0}(\bar{K}^{0}K^{{\ast}0})$ decays.
  The unavailable results are not listed. The other captions
  are the same as Table \ref{br}.}
  \label{cpps}
  \begin{ruledtabular}
  \begin{tabular}{c|cc|cc}
  $CP$ asymmetry  &
  \multicolumn{2}{c|}{$\bar{B}_{s}$ $\to$ $K^{\pm}K^{*\mp}$}
  & $\bar{B}_{s}$ $\to$ $\pi^{\pm}\rho^{\mp}$
  & $\bar{B}_{s}$ $\to$ $K^{0}\bar{K}^{{\ast}0}(\bar{K}^{0}K^{{\ast}0})$ \\
  parameters & This work & Cheng \cite{Cheng2}
       & This work & This work \\ \hline
  $C$        & $-4^{+0+2}_{-0-2}$   & $-8^{+4+15}_{-4-14}$   & 0 & $-0.44^{+0.01+0.06}_{-0.02-0.04}$ \\
  $S$        & $-9^{+0+6}_{-0-6}$   & $-5^{+1+13}_{-1-9}$    & $-74^{+2+0}_{-2-0}$ & $-0.1^{+0.0+0.0}_{-0.0-0.0}$\\
  $\Delta C$ & $ 8^{+1+14}_{-2-14}$ & $-3^{+12+46}_{-14-49}$ & 0 & $18^{+0+12}_{-0-14}$ \\
  $\Delta S$ & $28^{+1+8}_{-1-8}$   & $33^{+9+30}_{-10-48}$  & 0 & $37^{+0+6}_{-0-6}$ \\
  $A_{CP}$   & $24^{+1+1}_{-1-2}$   & $19^{+3+14}_{-4-11}$   & 0 & $-0.4^{+0.0+0.1}_{-0.0-0.1}$
  \end{tabular}
  \end{ruledtabular}
  \end{table}

  Based on above analyses, in order to get the theoretical
  predictions for $B_{s}$ $\to$ $PV$ decays,
  we assume that the endpoint parameters are universal
  for $B_{u,d,s}$ $\to$ $PV$ decays,
  and so we take
  \begin{equation}
  ({\rho}_{A,H}^i,{\phi}_{A,H}^i)=(3.08,-145^{\circ}),
  \quad
  ({\rho}_{A}^{f},{\phi}_{A}^{f})=(0.83,-36^{\circ}),
  \end{equation}
  gotten from $B_{u,d}$ $\to$ $PV$ decays \cite{chang4} as inputs.
  Now, we present the theoretical results for the
  observables of $B_{s}$ $\to$ $PV$ decays in
  Tables \ref{br}, \ref{cp} and \ref{cpps},
  in which the previous results in QCDF framework
  and in flavor symmetry (FS) framework are also
  summarized for comparison.
  It could be found that most of our predictions agree
  with the other theoretical results except for few
  tensions, which will be discussed in detail below.

  For the (color-suppressed) tree dominated $\bar{B}_{s}$
  $\to$ $\pi K^{*}$ and $\rho K$ decays, we find that:
  (1) even though our prediction for the branching fraction
  of observed $\bar{B}_{s}\to\pi^{-}K^{*+}$ decay agrees
  with experimental data given in Eq. (\ref{lhcbBspikstar})
  due to large theoretical uncertainties and experimental
  errors, the default value is still about twice larger
  than experimental central data.
  In fact, as noted in Ref. \cite{Cheng3}, all current
  theoretical results have such similar situation,
  even though these results are consistent with one
  another.
  So, more theoretical and experimental efforts are
  required to confirm or refute such possible mismatch.
  (2) The amplitudes of $\bar{B}_{s}$ $\to$ $\pi^{0}K^{*0}$
  and $K^{0}\rho^{0}$ decays are more sensitive to the
  HSS corrections related to large Wilson coefficient $C_1$.
  Due to a relative large $\rho_H$ is used in evaluations,
  our prediction for ${\cal B}(\bar{B}_{s}\to\pi^{0}K^{*0})$
  is larger than the results of  ``Cheng''~\cite{Cheng2}
  and ``S4'' \cite{Beneke2}, but much smaller than the
  one of ``FS'' \cite{Cheng3} where a large color-suppressed
  amplitude $C_{V}$ have similiar magnitude to, even larger than
  in Scheme C, the color-favored tree amplitude $T_{V}$.

  \begin{figure}[t]
  \subfigure[]{\includegraphics[width=5cm]{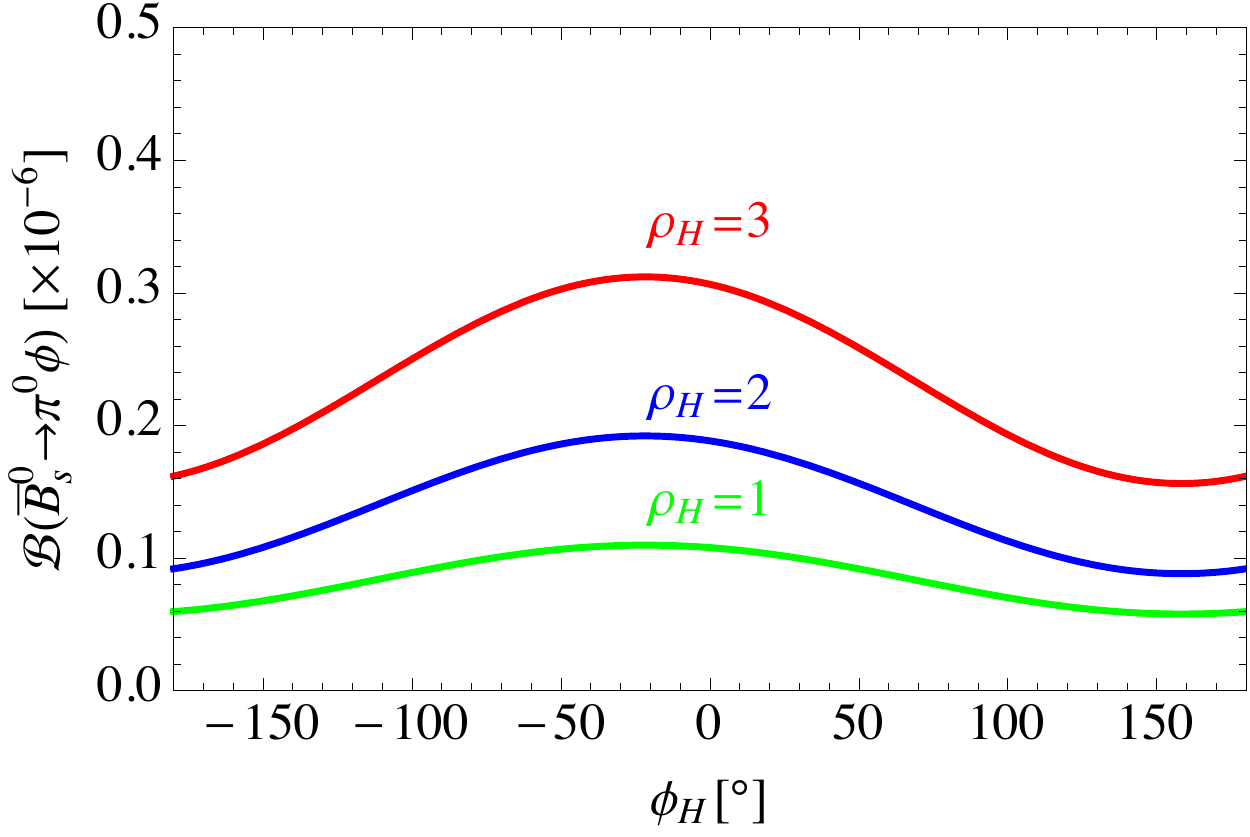}} \quad
  \subfigure[]{\includegraphics[width=5.2cm]{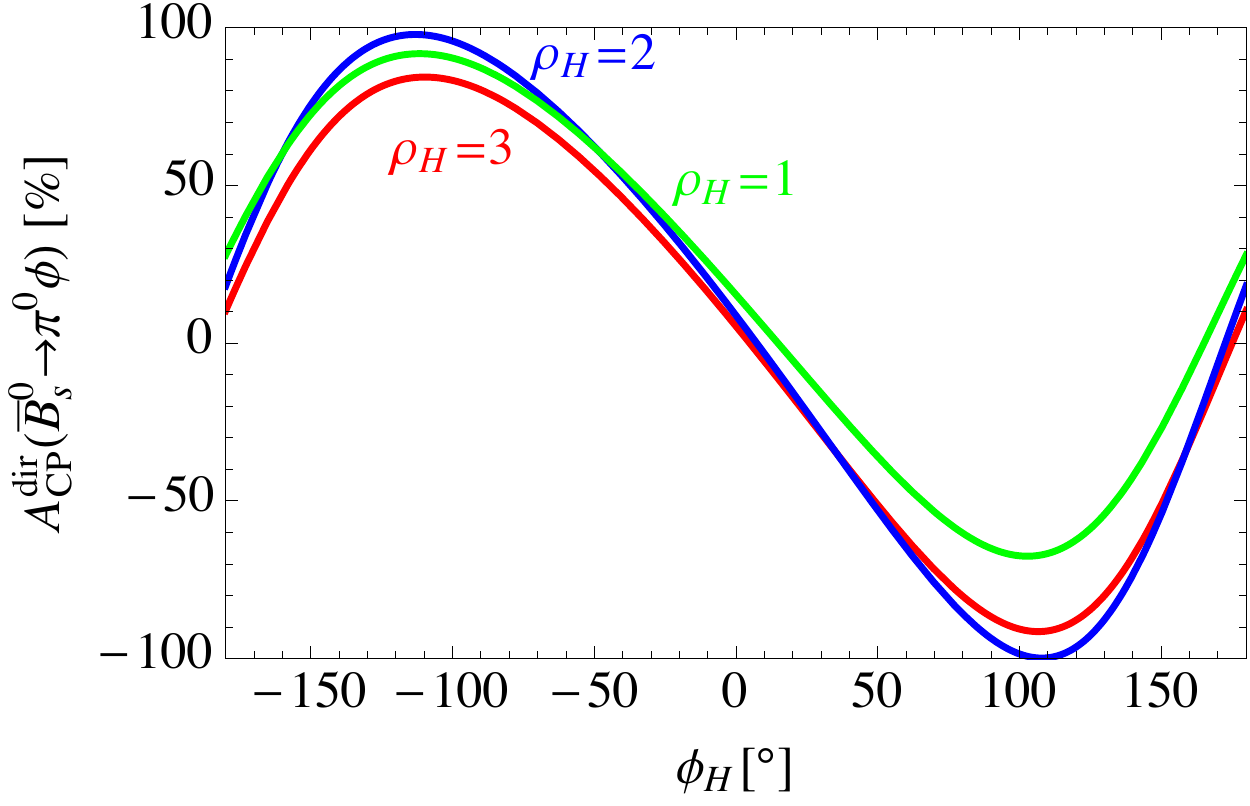}}\quad
  \subfigure[]{\includegraphics[width=5.2cm]{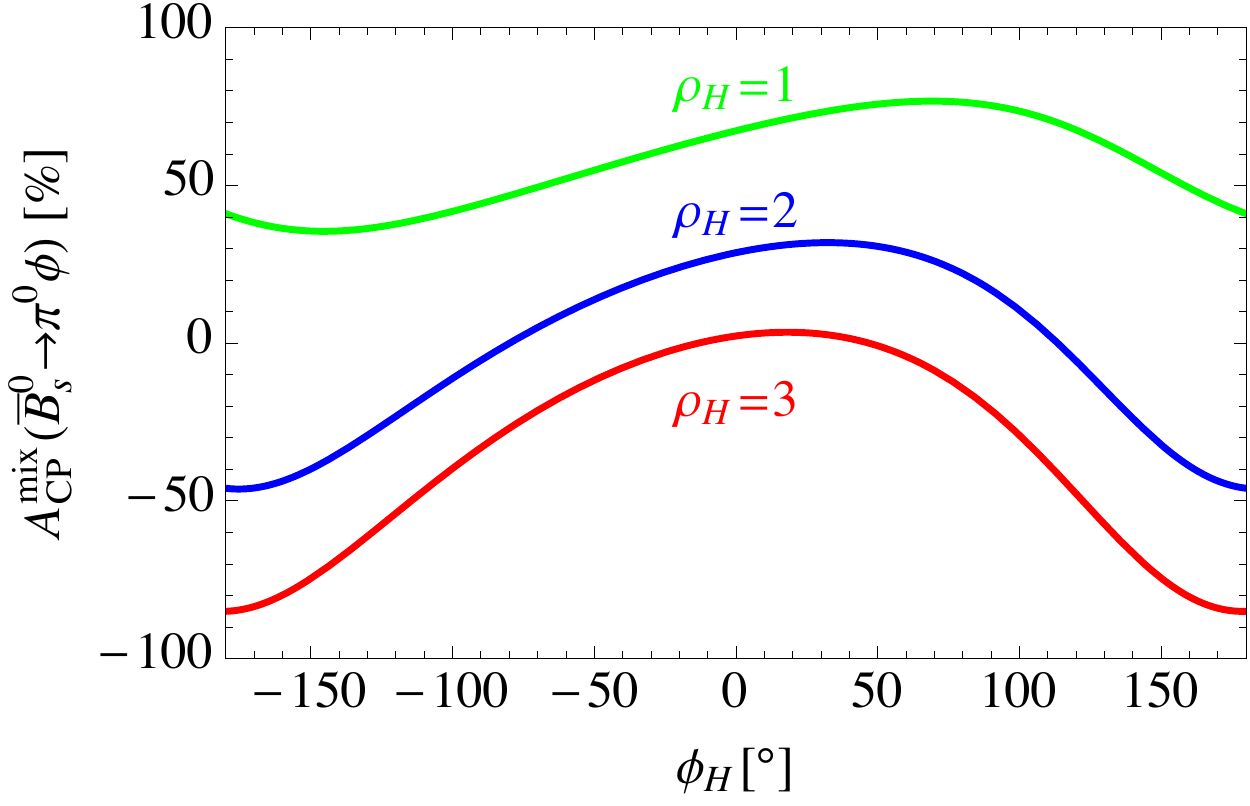}}
  \caption{The dependance of observables of $\bar{B}_{s}$ $\to$
  $\pi^{0}\phi$ decay on $\phi_H$.}
  \label{fig3}
  \end{figure}

   The color-suppressed tree and electroweak penguin dominated
   $\bar{B}_{s}$ $\to$ $\pi^{0}\phi$ decay is an important and
   interesting decay mode for exploring the HSS corrections,
   due to the fact that its amplitude,
   \begin{equation}
   {\cal A}({\bar{B}_{s}\to\pi^{0}\phi})\,
   {\propto}\, V_{ub}V_{us}^{\ast}\alpha_2+
    V_{tb}V_{ts}^{\ast}\,\frac{3}{2}\alpha_{3,EW}^p
   \label{apiphi},
   \end{equation}
   is sensitive to $\alpha_2$ related to possible large HSS
   corrections and irrelevant to the interference induced
   by WA corrections.
   Moreover, it plays an important role to judge the two
   direct ways through $\alpha_2$ or $\alpha_{3,EW}$
   respectively to resolve the so-called ``$\pi K$ CP puzzle"
   \cite{chang5,Hofer}.
   From Tables~\ref{br} and \ref{cp},
   it is found that our results for observables of
   $\bar{B}_{s}$ $\to$ $\pi^{0}\phi$ decay, especially
   the mixing-induced $CP$ asymmetry, are more or less
   different from the other ones.
   To clarify the reason, we present the dependence
   of observables on $\phi_H$ with different values of
   $\rho_H$ in Fig.~\ref{fig3}.
   It is found that the branching fraction is always
   about ${\cal O} (10^{-7})$, and hardly to be enhanced
   to ${\cal O} (10^{-6})$ level (predicted by ``FS'' \cite{Cheng3})
   by HSS corrections as Fig.\ref{fig3}(a) shows.
   From Fig.\ref{fig3}(b), it can be seen that the direct
   $CP$ asymmetry is much more sensitive to $\phi_H$ than
   the other observables, and so very suitable for exploring
   the possible strong phase in HSS.
   For $A_{CP}^{mix}(\bar{B}_{s} \to \pi^{0}\phi)$,
   one may note from Table~\ref{cp} that our prediction,
   even its sign, is significantly different from
   the others, which is an interesting finding and
   could be understood by the following reason.
   From Eq.(\ref{apiphi}), it can be found that
   whether $\alpha_2$ related to CKM factor
   $V_{ub}V_{us}^{*}$ or $\alpha_{3,EW}$ related
   to $V_{cb}V_{cs}^{*}$ (the part related to
   $V_{ub}V_{us}^{*}$ is negligible) dominates
   amplitude ${\cal A}_{\bar{B}_{s}\to\pi^{0}\phi}$
   is crucial for evaluating the mixing-induced $CP$
   asymmetry. As a result, as Fig.\ref{fig3}(c) shows,
   $A_{CP}^{mix}(\bar{B}_{s} \to \pi^{0}\phi)$ is
   very sensitive to $\rho_H$, and obviously negative
   when $\rho_H$ $\gtrsim$ 2.
   So, the future measurements on
   $A_{CP}^{mix}(\bar{B}_{s} \to \pi^{0}\phi)$
   will play an important role to probe the
   strength of HSS contribution.

   For the (electroweak) penguin dominated $\bar{B}_s$ decays,
   there are no significant difference between our predictions
   and the others within uncertainties.
   However, similar to the situation in $\bar{B}_{s}$ $\to$
   $\pi^{-}K^{*+}$ decay, the theoretical results for
   branching fraction of $\bar{B}_{s}$ $\to$ $K^{\pm}K^{*\mp}$
   decay listed in Table \ref{br} is significantly larger than
   the LHCb data in Eq.(\ref{lhcbBskkstar}).
   While, it is not a definite mismatch for now due to
   the large hadronic uncertainties.
   Besides of branching fraction, both $\bar{B}_{s}$ $\to$
   $\pi^{-}K^{*+}$ and $\bar{B}_{s}$ $\to$ $K^{0}\bar{K}^{{\ast}0}(\bar{K}^{0}K^{{\ast}0})$
   decays involve five non-zero $CP$ asymmetry observables
   listed in Table.\ref{cpps}, which would perform much
   stronger constraints on WA contributions if measured,
   especially on the possible strong phases therein.

   The pure annihilation $\bar{B}_{s}\to\pi\rho$ decays
   are principally suitable for probing the WA contributions
   without the interferences induced by HSS corrections.
   However, their branching fractions are stable at
   about ${\cal O}(10^{-8})$ level, and no significant $CP$
   asymmetries are found theoretically.
   So these decay modes are hardly to be measured very soon
   and possibly hard to provide useful information for
   the WA strong phases.

  \section{summary}
  \label{sec99}
   In summary, motivated by recent LHCb measurements on
   $\bar{B}_{s}$ $\to$ $\pi^{-}K^{*+}$ and $\bar{B}_{s}$ $\to$
   $K^{\pm}K^{*\mp}$ decays, we revisit the $\bar{B}_{s}$ $\to$
   $PV$ decays within the QCD factorization framework,
   and analysis the effects of HSS and WA corrections
   related to end-point parameters in detail.
   Our main findings and conclusions could be summarized
   as follows:
  \begin{itemize}
  \item
  By $\chi^2$ fitting, it is found that the WA parameters
  in $ \bar{B}_{s}$ $\to$ $PV$ decays could not be well bounded
  due to large theoretical uncertainties and rough experimental
  measurements, and the universalities of their values in
  $B_s$ and $B_{u,d}$ system are still allowed.
  Assuming the values of (${\rho}_{A}^{i,f}$, ${\phi}_{A}^{i,f}$)
  are universal for $B_s$ and $B_{u,d}$ decays and
  (${\rho}_{H}$, ${\phi}_{H}$) $=$ (${\rho}_{A}^{i}$, ${\phi}_{A}^{i}$),
  our theoretical predictions are presented in
  Tables \ref{br}, \ref{cp} and \ref{cpps},
  which will be tested in the near future.
  \item
  For the branching fractions of observed $\bar{B}_{s}$ $\to$
  $\pi^{-}K^{*+}$ and $K^{\pm}K^{*\mp}$ decays, even though
  the current theoretical results agree with each other and
  are consistent with data within errors,
  their default results are significantly much larger than
  the central values of data.
  The refined measurements on such two observables will
  perform strong constraints on (${\rho}_{H}$, ${\phi}_{H}$)
  and (${\rho}_{A}^{f}$, ${\phi}_{A}^{f}$), respectively,
  and further to check such possible mismatches.
  \item
  A detailed analysis for the other observables and
  decays modes are perform, and some interesting
  findings are presented in text. Especially,
  for instance, without the interference induced
  by annihilation corrections, $\bar{B}_{s}$ $\to$
  $\pi^{0}\phi$ decay plays an important role to
  exploring the HSS contributions, where the size
  and strong phase could be clearly determined
  by direct and mixing-induced CP asymmetries,
  respectively.
  In addition, the pure annihilation  $\bar{B}_{s}$
  $\to$ $\pi\rho$ decays will be suitable for probing
  the WA contributions, although their branching
  ratios are very small.
  \end{itemize}

  Moreover, more $B_s$ meson decay modes are urgently
  expected to be measured soon by LHCb and upgrading
  Belle II in the near future,
  which will exhibit exact picture of WA and HSS
  contributions in $B_s$ decays.

  \section*{Acknowledgments}
  The work is supported by the National Natural Science Foundation of China
  (Grant Nos. 11475055, 11275057 and U1232101).
  Q. Chang is also supported by the Foundation for the Author of National
  Excellent Doctoral Dissertation of P. R. China (Grant No. 201317)
  and the Program for Science and Technology Innovation Talents in
  Universities of Henan Province (Grant No. 14HASTIT036).

  
 \end{document}